\documentclass[12pt]{article}
\usepackage{epsf}
\def\be{\begin{equation}}
\def\ee{\end{equation}}
\def\bea{\begin{eqnarray}}
\def\eea{\end{eqnarray}}

\begin{document}
\begin{titlepage}
\begin{center}
{\Large \bf William I. Fine Theoretical Physics Institute \\
University of Minnesota \\}
\end{center}
\vspace{0.2in}
\begin{flushright}
FTPI-MINN-16/10 \\
UMN-TH-3521/16 \\
March 2016 \\
\end{flushright}
\vspace{0.3in}
\begin{center}
{\Large \bf $\Upsilon(6S)$ and triangle singularity in $e^+e^- \to B_1(5721) \bar B \to Z_b(10610) \, \pi$ 
\\}
\vspace{0.2in}
{\bf  A.E. Bondar$^{1,2}$ and M.B. Voloshin$^{3,4,5}$  \\ }
$^1$Budker Institute of Nuclear Physics, 630090
Novosibirsk, Russia \\
$^2$Novosibirsk State University,
630090 Novosibirsk, Russia \\
$^3$William I. Fine Theoretical Physics Institute, University of
Minnesota,\\ Minneapolis, MN 55455, USA \\
$^4$School of Physics and Astronomy, University of Minnesota, Minneapolis, MN 55455, USA \\ 
$^5$Institute of Theoretical and Experimental Physics, 117218 Moscow,  Russia
\\[0.2in]

\end{center}

\vspace{0.2in}

\begin{abstract}
We discuss the possibility that production of final states with bottomonium and light mesons at the peak $\Upsilon(6S)$ in the $e^+e^-$ annihilation at approximately 11.00\,GeV is in  fact due to a triangular singularity at the threshold of the heavy meson pair production $B_1(5721) \bar B + c.c.$ through the process
$e^+e^- \to B_1(5721) \bar B \to Z_b(10610) \, \pi$. The presence of the hidden-bottom resonance $Z_b(10610)$ then explains the observed enhanced production of the final channels with both ortho- and para- bottomonium states, $\Upsilon(nS) \pi \pi$ and $h_b(kP) \pi \pi$. The discussed mechanism also predicts a distinct pattern for production of hidden-bottom states at the $\Upsilon(6S)$ energy that can be tested by experiment. 

\end{abstract}
\end{titlepage}

Recent experimental studies~\cite{belle1501,belle1508} of the $e^+e^-$ annihilation in the energy range of the $\Upsilon(6S)$ peak at approximately 11.00\,GeV have found certain differences in the patterns of final states from those in the lower mass peak $\Upsilon(5S)$. In particular, at both peaks there is a measurable production of the decay channels with both ortho- and para- states of bottomonium, $\Upsilon(nS) \pi \pi$ ($n=1,2,3$) and $h_b(kP) \pi \pi$ $(k=1,2)$, and in both peaks the latter decays, violating the Heavy Quark Spin Symmetry (HQSS), appear to be associated with the $Z_b(10610)$ and/or $Z_b(10650)$ resonances~\cite{bellez} by the mechanism described in Ref.~\cite{bgmmv}. However there appears to be a difference in the behavior for the HQSS allowed channels $\Upsilon(nS) \pi \pi$. Namely, on one hand, in the $\Upsilon(5S)$ peak there is a significant fraction of the yield in these channels outside of the $Z_b$ resonances, unlike the production of the $h_b(kP) \pi \pi$ channels which goes exclusively through the $Z_b$ resonances.  On the other hand, the ratio of the yield of  $\Upsilon(nS) \pi \pi$  and $h_b(kP) \pi \pi$ across the $\Upsilon(6S)$ is smaller than in $\Upsilon(5S)$ and suggests~\cite{belle1501,belle1508} that in the $\Upsilon(nS) \pi \pi$ channels from $\Upsilon(6S)$ there is also very little or no non-resonant production not associated with the $Z_b$ intermediate states. Motivated by this observation, we discuss here the possibility that the decays of $\Upsilon(6S)$ into final states with bottomonium are of a different origin than those of $\Upsilon(5S)$.  Namely, these production channels are boosted by a `threshold bump' due to the so-called triangle singularity in the process $e^+e^- \to B_1(5721) \bar B \to Z_b(10610) \, \pi$ due to the decay $B_1 \to B^* \pi$, and the $B^* \bar B$ pair forming the $Z_b(10610)$ resonance as shown in Fig.~1.  The triangle singularity arises when all three particles in the loop are on the mass shell, and the spread of the bump is a result of `smearing' of the `mass shell' due to the widths of the resonances. Possible existence of such threshold bumps in hadronic processes was suggested long ago~\cite{lt} and more recently a similar picture in the hidden-charm sector was discussed~\cite{ll} in connection with the structure $Y(4260)$ (and it was also suggested~\cite{lv} that a similar bump may occur for hidden bottom at 11\,GeV). The presented here interpretation of the bottomonium production at $\Upsilon(6S)$ implies the following distinct features that should be observable in $e^+e^-$ annihilation and that can be tested experimentally in the existing and/or future data: 
\renewcommand{\theenumi}{\it \roman{enumi}}%
\begin{enumerate}
  \item The production of final states with bottomonium at $\Upsilon(6S)$ proceed through the $Z_b(10610)$ resonance with no non-resonant background.
  \item Only the $Z_b(10610)$ is present in the production channels, but not the $Z_b(10650)$. (The current data~\cite{belle1508} could not resolve the two $Z_b$ resonances in the $\Upsilon(6S)$ peak.)
  \item There should be a detectable production of $B_1(5721) \bar B + c.c.$ heavy meson pairs in the threshold region. In particular, this should contribute to the yield of the final channel $(B^* \bar B + c.c.) \, \pi$, but not $B^* \bar B^* \pi$.
  \item The sub dominant decay of the $B_1$ meson, $B_1 \to B \pi \pi$, may provide, through a similar mechanism, a gateway to studies of the expected~\cite{mv12} at the $B \bar B$ threshold resonance $W_{b0}$ with quantum numbers $I^G(J^P)=1^-(0^+)$.
  \item Additionally, there may be another similar bump at the c.m. energy around 11.06\,GeV, near the threshold of $B_1 \bar B^*$ and possibly $B_2 \bar B^*$, where the production of channels with bottomonium may proceed through a mixture of the $Z_b(10610)$ and $Z_b(10650)$ resonances. (At present there is  no appropriate data at $e^+e^-$ energies above 11.02\,GeV.)
\end{enumerate}

\begin{figure}[ht]
\begin{center}
 \leavevmode
    \epsfxsize=12cm
    \epsfbox{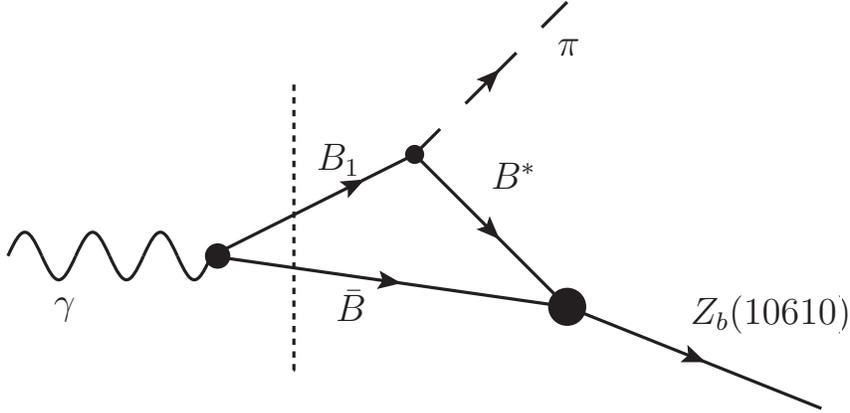}
    \caption{The graph for the mechanism generating the triangle singularity in the process $e^+e^- \to B_1(5721) \bar B \to Z_b(10610) \, \pi$. The thin dashed line shows the unitarity cut.}
\end{center}
\end{figure}
 
We emphasize that the discussed effect of the triangle singularity arises specifically in the $Z_b(10610) \pi$ channel on top of any other features of the hidden bottom production that may be present at the energies in the range of $\Upsilon(6S)$ in other channels, e.g. due to a near-threshold enhancement of the $B_1 \bar B + c.c.$ channel.  Clearly, a presence or absence of such features can be studied separately in those other channels.

It has to be noted, however, that the discussed here picture has a known caveat. Namely, it has to be assumed that there is a production of  the meson pairs $B_1(5721) \bar B + c.c.$  in the $S$ wave, since it is highly unlikely that a $D$ wave production, heavily suppressed near the threshold by the phase space factor, would result in a threshold bump. If the $B_1(5721)$ is treated as the lower mass state in the so-called ${3 \over 2}^+$ doublet of excited bottom mesons, $B_1(5721)$ and $B_2(5747)$, where the light antiquark is in the $P$ wave state with quantum numbers ${3 \over 2}^+$, the $S$-wave amplitude of pair production in the $e^+e^-$ annihilation of any of these mesons together with the corresponding ground state ${1 \over 2}^-$ meson ($B$ or $B^*$) is forbidden by HQSS~\cite{lv}. A similar difficulty also applies to the  models of the hidden-charm structure $Y(4260)$ as a $D_1(2420) \bar D$ molecule~\cite{ding,lwdz,whz,cwghmz} or as a threshold bump~\cite{ll} due to triangular singularity. Clearly, a resolution of this difficulty requires a violation of HQSS. One source of such symmetry breaking can be a mixing between the $J^P=1^+$ meson from the ${3 \over 2}^+$ doublet with the axial meson from the ${1 \over 2}^+$ doublet, where the light antiquark is in the ${1 \over 2}^+$ state. Normally the $J^P=0^+$ and $1^+$ heavy mesons in the latter doublet are expected to be broad due to their $S$-wave decay into respectively $B \pi$ and $B^*\pi$ (in the $B$ sector, for definiteness). This is different from the case of the mesons in the ${3 \over 2}^+$ doublet in that the latter mesons decay into $D$ wave and thus have smaller widths. Based on these decay properties, there is some indication of a mixing between the axial mesons from two excited doublets. Indeed, the LHCb experiment recently measured~\cite{lhcb} the widths of the $B_1$ and $B_2$ with uncertainty of about $1 \div 2$\,MeV and a similar difference between the neutral and charged mesons. We use here as rounded representative values $\Gamma(B_1) = 30\,$MeV and $\Gamma(B_2)=24\,$MeV~\footnote{We also use similarly rounded values of the measured~\cite{lhcb} masses $M(B_1)=5726\,$MeV and $M(B_2)=5738\,$MeV, so that the `nominal' position of the threshold for $B_1 \bar B + c.c.$ pairs is estimated as 11006\,MeV.}. For a $B_1$ meson being a pure (unmixed) component of the ${3 \over 2}^+$ doublet the HQSS relation for its width of decay $B_1 \to B^* \pi$ in terms of $\Gamma(B_2)$ reads as
\be
\Gamma(B_1 \to B^* \pi) = {5 \, k^5  \over 2 \, k_0^5 + 3 \, k_1^5} \, \Gamma(B_2) \approx 16\,{\rm MeV}~,
\label{g12}
\ee
where $k \approx 362\,$MeV is the pion momentum in the decay $B_1 \to B^* \pi$, and $k_0 \approx 418\,$MeV and $k_1 \approx 374\,$Mev are the respective pion momenta in the decays $B_2 \to B \pi$ and $B_2 \to B^* \pi$. The deficit of about 14\,MeV in comparison with the measured total width of $B_1$ can be attributed to an enhancement due to presence of an $S$ wave in the decay arising from a mixing with the $J^P=1^+$ meson from the  ${1 \over 2}^+$ doublet, although this estimate can be somewhat reduced due to existence of the decay $B_1 \to B \pi \pi$, which has not been observed, but is expected based on the similar decay of charmed mesons $D_1 \to D \pi \pi$~\cite{belle05}. The branching fraction for the latter decay is unknown but is generally assumed to be small~\footnote{It can be also noted that a similar deficit of approximately 15\,MeV can be deduced for the total width of the charmed $D_1(2420)$ meson as compared to a HQSS calculation from the width of $D_2(2460)$}. It is not clear at present how this indication of the mixing should be interpreted quantitatively given large uncertainties in the current knowledge of the parameters of the heavy mesons in the ${1 \over 2}^+$ doublet.

Admittedly, at present we can offer no explanation for an $S$-wave production of the heavy meson pairs $B_1 \bar B + c.c.$ in the $e^+e^-$ annihilation. However, assuming that such production takes place, we can estimate the significance of the effect of the triangle singularity by 
evaluating the absorptive part of the amplitude generated by the mechanism of Fig.~1~\footnote{The calculation described here is in fact similar to the one in Ref.~\cite{dv} for the process $e^+e^- \to D^* \bar D^* \to X(3872) \, \gamma$ near the $D^* \bar D^*$ threshold.}. For this calculation one needs the amplitude for the production of $B_1 \bar B$ ($\bar B_1 B$) by the electromagnetic current and the amplitude for the process $B_1 \bar B$ ($\bar B_1 B$) $\to Z_b(10610) \, \pi$. The assumed $S$-wave part of the electromagnetic vertex can be written in terms of an effective Lagrangian for the interaction of the current $\vec j$ of the electrons with the heavy meson pairs near the $B_1 \bar B$ threshold
\be
L_{B_1 B \, \gamma}= {C \over \sqrt{2}} \, j_i \, \left ( B_{1i}^+ B^- - B_{1i}^- B^+ + B_{1i}^0 \bar B^0 - \bar B_{1i}^0 B^0 \right )~,
\label{lbbg} 
\ee
where $B_{1i}$ stands for the polarization amplitude of the $B_1$ meson, and a nonrelativistic normalization of the wave functions for heavy mesons is assumed throughout the present discussion. The overall constant $C$  generally depends on  the c.m. energy $E=\sqrt{s}$ and this dependence may or may not contain additional near-threshold features in the discussed channel.

The effective Lagrangian for the coupling between the $B_1$ mesons and the $B^* \pi$ decay channels can be generally written as
\be
L_{B^* \pi \, B_1} = {g_0 \over \sqrt{2}} \left ( B^{* \dagger}_i \tau^a B_{1i} \right ) \, \partial_0 \pi^a +  {g_2 \over \sqrt{2}} \left ( B^{* \dagger}_i \tau^a B_{1j} \right ) \, \left ( \partial_i \partial_j - {1 \over 3} \delta_{ij} \vec \partial^{\,2} \right ) \pi^a +h.c.
\label{lbbpi}
\ee
where $a$ is the isotopic triplet index, and the time derivative in the first term is mandated by the chiral algebra requirement that the amplitude goes to zero at zero four-momentum of the pion. The constants $g_0$ and $g_2$ describe the $S-$ and $D-$wave amplitudes in the decay $B_1 \to B^* \pi$. The rate of the decay is given, in terms of these constants, as
\be
\Gamma(B_1 \to B^* \pi^+) = 2 \, \Gamma(B_1 \to B^* \pi^0) = |g_0|^2 \, {\omega^2 \, k \over 2 \pi} + |g_2|^2 \, {k^5 \over 9 \pi}
\label{gbbpi}
\ee
with $k = |\vec k|$ and $\omega$ being the momentum and the energy of the emitted pion. [In the subsequent treatment we neglect the small variation of $k$ across the width of the $B_1$ resonance and across that of the $\Upsilon(6S)$ peak and set it at its `nominal' value as in Eq.(\ref{g12})].

We consider the $Z_b(10610)$ resonance as a shallow $S$-wave bound state of heavy mesons $B^* \bar B - \bar B^* B$ with the binding energy $E_b =-\varepsilon$. At  small $\varepsilon$ the mesons in the bound state move at characteristic distances set by the scale $a=\kappa^{-1}$ with $\kappa$, the characteristic momentum of each of the mesons in the bound state, being  given by $\kappa = \sqrt{M \, \varepsilon} \approx 73\,{\rm MeV} \, \sqrt{\varepsilon/1\,{\rm MeV}}$ where $M \approx 5300\,$MeV is standing for twice the reduced mass in a system of $B^*$ and $B$. Since the energy $\varepsilon$ is in the ballpark of 1\,MeV, the mesons dominantly move well beyond the range of strong interaction, and their wave function can be approximated (in the momentum space) as
\be
\phi(\vec q)= {\sqrt{8 \pi \, \kappa} \over \vec q^{\,2} + \kappa^2}~.
\label{wf}
\ee
It should be noted that at large momenta, comparable to the strong interaction scale $\Lambda$ (i.e. at short distances $r < \Lambda^{-1}$), this  expression is not applicable and should be modified. The calculation discussed here is strictly in the leading order at $\Lambda \to \infty$, and any effects of a finite spatial range of the strong interaction are beyond the accuracy of our estimates. 

The wave function (\ref{wf}) can be used to find an expression for the amplitude of the conversion of the state of the $B_1 \bar B + c.c.$ pairs produced by the electromagnetic current [Eq.(\ref{lbbg})] into the final state $Z_b(10610) \pi$ resulting from the decay $B_1 \to B^* \pi$ ($\bar B_1 \to \bar B^* \pi$) and a subsequent coalescence of the bottom vector and pseudoscalar mesons into $Z_b(10610)$. Considering for definiteness the final channel with specific charges: $Z_b^- \, \pi^+$, and taking into account the molecular structure of the $Z_b^-(10610)$ in terms of the mesons~\cite{bgmmv}: $Z_b(10610) \sim (B^{*0} B^- - B^{*-} B^0)/\sqrt{2}$, one can write the amplitude of this conversion as
\bea
&&A[(B_1 \bar B - c.c.) \to  Z_b^-(10610) \, \pi^+] = \nonumber \\
&&{1 \over \sqrt{2}} \left \langle Z_b^- \, \pi^+ \left | L_{B^* \pi \, B_1} \right | B_{1}^+(\vec p, \vec \epsilon_B) \,  B^-(-\vec p) - \bar B_{1}^0(\vec p, \vec \epsilon_B) B^0(- \vec p) \right \rangle = \nonumber \\
&&\left [ -i \, g_0 \omega (\vec \epsilon_B \cdot \vec \epsilon^{ \, *}_Z) - g_2 \left ( k_i k_j - {1 \over 3} \, \delta_{ij} \, k^2 \right ) \, \epsilon_{Bi} \epsilon_{Zj}^* \right ] \, \phi\left ( \vec p - {1 \over 2} \, \vec k \right )~,
\label{abbzp}
\eea
where $- \vec p$ ($\vec p$) is the c.m. momentum of the axial (pseudoscalar) heavy meson, $\vec \epsilon_B$ is the polarization amplitude of the axial meson (that defines the total polarization amplitude of the heavy meson pair produced in the $S$ wave), and $\vec \epsilon_Z$ is the polarization amplitude of the $Z_b$ resonance.

In the present calculation we take into account the finite width of the $B_1$ meson (but not the smaller width of the $Z_b$ resonance). This is done in the Breit-Wigner approximation by considering the (invariant) mass $\mu$ of the resonance being spread around the `nominal' mass $M(B_1)$ with the density 
\be 
-{1 \over \pi} {\rm Im} D_{BW}(\mu) = {1 \over \pi} {\Gamma(B_1)/2 \over [\mu - M(B_1)]^2 +  \Gamma^2(B_1)/4}
\label{sd}
\ee
(which density, naturally, becomes $\delta[\mu - M(B_1)]$ in the limit of vanishing resonance width $\Gamma$). 

Using the equations (\ref{lbbg}) and (\ref{abbzp}) one can readily write the expression for the absorptive part of the amplitude $A[e^+e^- \to Z_b^-(10610) \, \pi^+]$ corresponding to the unitarity cut shown in Fig.~1 in the form
\bea
&&A_{abs}[e^+e^- \to Z_b^-(10610) \, \pi^+] = C \left [ i \, g_0 \omega \, (\vec j \cdot \vec \epsilon^{ \, *}_Z) + g_2 \left ( k_i k_l - {1 \over 3} \, \delta_{i l} \, k^2 \right ) \, j_i \epsilon_{Z l}^* \right ] \times \nonumber \\
&&{1 \over 2} \int  \phi\left ( \vec p - {1 \over 2} \, \vec k \right ) \left [ {\rm Im} D_{BW} (\mu) \right ] \, 2 \pi \, \delta \left [ E - M(B)- \mu - {p^2 \over M_1} \right ] \, {d^3 p \over (2 \pi)^3} \, { d \mu \over  \pi}~,
\label{agzp}
\eea 
where $E= \sqrt{s}$ is the total c.m. energy, and $M_1 \approx 5495\,$MeV is twice the reduced mass in the system $B_1 \bar B$ (a small variation of this reduced mass across the width of $B_1$ is neglected).

One can readily notice that the only angular dependence in the integrand in Eq.(\ref{agzp}) is that of $\phi (\vec p - \vec k/2)$ on the angle $\theta$ between the momenta $\vec p$ and $\vec k$. Thus the wave function from Eq.(\ref{wf}) can be replaced by its angular average, depending only on the absolute values $p$ and $k$:
\be
\phi\left ( \vec p - {1 \over 2} \, \vec k \right ) \to {1 \over 2} \, \int \, \phi\left ( \vec p - {1 \over 2} \, \vec k \right ) \, d \cos \theta = {\sqrt{2 \pi \kappa} \over  p \, k} \, L(p)~,
\label{avg}
\ee
with the dimensionless factor $L(p)$ given by
\be
L(p)= \log { (p+k/2)^2 + \kappa^2 \over (p- k/2)^2 + \kappa^2}~.
\label{defl}
\ee
After this simplification the amplitude in Eq.(\ref{agzp}) can be written in the form
\be
A_{abs}[e^+e^- \to Z_b^-(10610) \, \pi^+] = C \left [ i \, g_0 \omega (\vec j \cdot \vec \epsilon^{ \, *}_Z) + g_2 \left ( k_i k_l - {1 \over 3} \, \delta_{i l} \, k^2 \right ) \, j_i \epsilon_{Z l}^* \right ] \, {M_1 \, \sqrt{\kappa} \over \sqrt{8 \pi} \, k} \, \Phi(E)~,
\label{aphi}
\ee
where $\Phi(E)$ is dimensionless and reads as
\be
\Phi(E) = \int \, L(p) \, \left [ {\rm Im} D_{BW} (\mu) \right ] \, {d \mu \over  \pi}~,
\label{dphi}
\ee
with $p$ being a function of $\mu$ determined by the energy conservation: $p(\mu)=\sqrt{M_1 \, [E - M(B) - \mu]}$.

When calculating the cross section generated by the amplitude (\ref{aphi}) it is helpful to notice that the pion emission part, described by the constants $g_0$ and $g_2$, factorizes out, so that the integration of the square of this part over the phase space of the pion reduces to the width of the decay $B_1 \to B^* \pi$ as in Eq.(\ref{gbbpi}). Thus, taking into account all the charge combinations, the cross section for the process $e^+e^- \to Z_b(10610) \, \pi$, generated by the absorptive part of the graph in Fig.~1, can be expressed as
\be
\sigma[e^+e^- \to Z_b(10610) \, \pi] = C_1 \, {M_1^2 \, \kappa \, \Gamma(B_1 \to B^* \, \pi)  \over 8 \pi \, k^2} \, \Phi^2(E)~,  
\label{szpi}
\ee
where the constant $C_1$ is proportional to $|C|^2$ and can be related to the cross section of production of the meson pairs $B_1 \bar B + c.c.$ in continuum as
\be
\sigma(e^+ e^- \to B_1 \bar B + c.c.) = C_1 \, {M_1 \, P(E) \over 2 \pi}~.
\label{sbb}
\ee
Here the averaged momentum $P(E)$ takes into account the Breit-Wigner spread of the invariant mass of the $B_1$ meson:
\be
P(E) = - \int \, p(\mu) \, \left [ {\rm Im} D_{BW} (\mu) \right ] \, {d \mu \over \pi}~.
\label{pavg}
\ee

We illustrate the effect of the discussed triangle singularity as estimated from Eq.(\ref{szpi}) in Fig.~2 with the plots of the shape function $\Phi^2(E)$ and in Fig.~3 with the plots of the ratio of the cross sections $R_Z = \sigma[e^+e^- \to Z_b(10610) \, \pi]/\sigma(e^+ e^- \to B_1 \bar B + c.c.)$. In the latter plots we assume that $\Gamma(B_1 \to B^* \pi) \approx \Gamma(B_1) = 30\,$MeV.

\begin{figure}[ht]
\begin{center}
 \leavevmode
    \epsfxsize=10cm
    \epsfbox{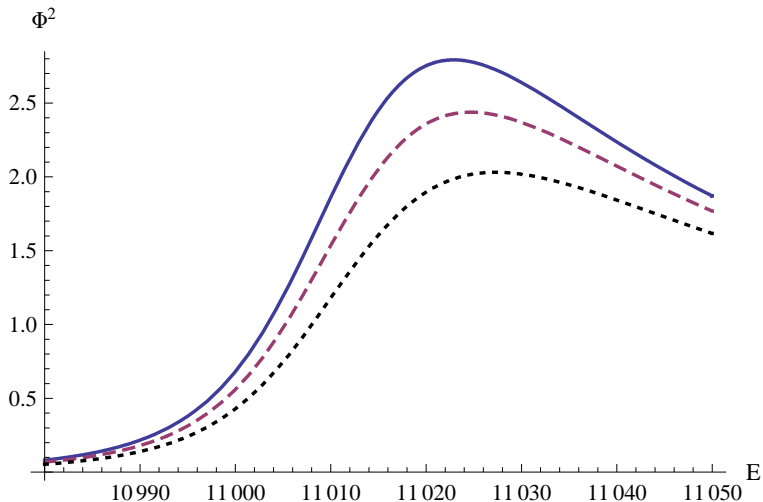}
    \caption{The shape function $\Phi^2(E)$ for excitation of the final channel $Z_b(1050) \, \pi$ generated by the triangle singularity [Eq.(\ref{szpi})] at representative values of the binding energy $\varepsilon$ of $B^* \bar B$ in $Z_b$: 0.5\,MeV (solid), 1\,MeV (dashed), 2\,MeV (dotted).}
\end{center}
\end{figure}
\begin{figure}[ht]
\begin{center}
 \leavevmode
    \epsfxsize=10cm
    \epsfbox{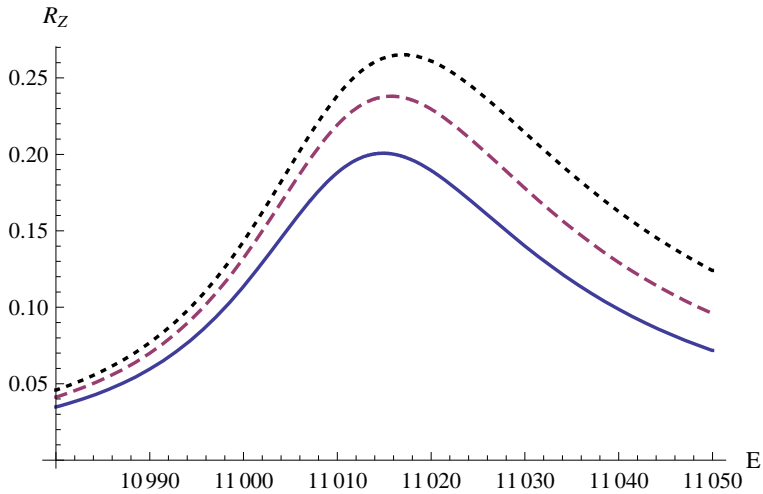}
    \caption{The ratio of the cross sections $R_Z= \sigma[e^+e^- \to Z_b(10610) \, \pi]/\sigma(e^+ e^- \to B_1 \bar B + c.c.)$ [Eqs.(\ref{szpi}) and (\ref{sbb})] at representative values of the binding energy $\varepsilon$ of $B^* \bar B$ in $Z_b(10610)$: 0.5\,MeV (solid), 1\,MeV (dashed), 2\,MeV (dotted).}
\end{center}
\end{figure}

The estimate of the threshold enhancement in Eq.(\ref{szpi}) is based on the evaluation of the absorptive part of the amplitude in Eq.(\ref{aphi}). The dispersive part resulting from the triangle graph of Fig.~1 generally does not display such enhancement and is a smooth function of energy. Moreover, a calculation of the latter part requires knowledge of unknown vertex form factors for off-shell mesons and also 
depends on contribution of other intermediate channels. An indirect indication of a small value of the smooth dispersive part is provided by the experimental observation~\cite{belle1501,belle1508} of very little, if any, background under the $\Upsilon(6S)$ peak for production of final states with bottomonium. This, in particular, motivates our conclusion that all the production of such final states in the peak proceeds due to the triangle singularity and hence through the $Z_b(10610)$ resonance. The fact that only this resonance gives contribution, and not the $Z_b(10650)$, simply follows from that only the decay $B_1 \to B^* \pi$ is possible with a single pion, so that only a threshold molecular state made from $B^*$ and $\bar B$ can be formed, i.e. the lower $Z_b(10610)$ resonance. 

The production of the heavier $Z_b(10650)$ resonance, considered to be a threshold molecular state of $B^* \bar B^*$, can be expected through the same triangle singularity mechanism near the threshold of $B_1 \bar B^*$ at approximately 11.052\,GeV, where no suitable data on the $e^+ e^-$ annihilation are currently available. It should be noted, however, that the mass of the tensor $B_2$ meson is measured~\cite{lhcb} to be only (10 - 15)\,MeV heavier than that of the $B_1$, so that the separation between the thresholds for the pairs $B_1 \bar B^*$ and $B_2 \bar B^*$ is less than the spread due to the widths of $B_1$ and $B_2$. Since the mechanism for the assumed HQSS-breaking $S$-wave production of the pairs $B_1 \bar B$ is currently unknown, it is not clear whether a similar threshold production of $B_2 \bar B^* + c.c.$ takes place. In particular, an $S$ wave in the latter channel should not be present if in the former channel the threshold production is due to the discussed mixing of axial mesons from the ${3 \over 2}^+$ and ${1 \over 2}^+$ doublets. However, if the mechanism of HQSS violation is different and both $B_1 \bar B^*$ and $B_2 \bar B^*$ are produced in $S$ wave near threshold, a more complicated structure can exist near 11.06\,GeV due to the presence and interference between these two channels in their decay products. Furthermore, the tensor $B_2$ meson decays into both $B^* \pi$ and $B \pi$. The $B$ meson from the latter decay can coalesce with the $\bar B^*$ meson into the $Z_b(10610)$ resonance through the mechanism similar to that in Fig.~1, so that in this case there should be a presence of this resonance along with the $Z_b(10650)$. We can only hope at this point that an experimental study of potentially quite intricate properties of a possible structure near 11.06\,GeV may shed light on the presently unknown details of the heavy meson dynamics.

The underlying process for the considered here yield of $Z_b \pi$ is the production of heavy meson pairs $B_1 \bar B + c.c.$. Therefore, for the discussed mechanism to work there should be a measurable cross section for the latter channel. Hopefully the yield of the $B_1$ mesons can be probed by either their dominant decay into $B^* \pi$ or the sub dominant mode $B_1 \to B \pi \pi$.
The former decay should contribute to the production of the final channel $(B^* \bar B + c.c.) \, \pi$ with the heavy meson pair not originating from the $Z_b(10610)$ resonance. Thus it should be expected that the ratio of the yield of $h_b(kP) \pi \pi$ to that of $(B^* \bar B + c.c.) \, \pi$ should be smaller in the $\Upsilon(6S)$ peak than in the $\Upsilon(5S)$ resonance where both final channels go through the $Z_b(10610)$ resonance~\cite{belle1512}. Moreover, in a large, if not dominant, fraction of the decays $B_1 \to B^* \pi$ the pion is emitted in the $D$ wave. A presence of a $D$-wave pion both in the channel $(B^* \bar B + c.c.) \, \pi$ and in the channels associated with the discussed here process $B_1 \bar B \to Z_b(10610) \pi$ can be established by an angular analysis with future data. In addition one should also expect a strong suppression of the final channel $B^* \bar B^* \pi$ in comparison with $(B^* \bar B + c.c.) \, \pi$.

 The decay $B_1 \to B \pi \pi$ also raises a tantalizing possibility of studying threshold behavior of $B \bar B$ pairs~\footnote{The decay $B_1 \to B \pi \pi$ has not been observed. Based on similar decays~\cite{pdg} $D_1(2420) \to D \pi \pi$ and also $K_1(1270) \to K \pi \pi$, this decay should contribute a sub-dominant, but still a sizable fraction of the total width of $B_1$, with a significant part of the yield in the channel with an isovector dipion.  One can hope that this decay can be studied in the LHCb experiment}.  In particular, if this decay is contributed by emission of the dipion in the isovector state, the $B \bar B$ in the recoil to dipion is in the $I^G=1^-$ isotopic state. A threshold resonance $W_{b0}$ with these quantum numbers and $J^P=0^+$ is expected~\cite{mv12} from an HQSS-based relation to the $Z_b$ resonances. The cross section for the process $e^+e^- \to W_{b0} \pi \pi$ should then be enhanced at $\Upsilon(6S)$ due to the triangle singularity of the same type as shown in Fig.~1 with the single pion emission being replaced by that of the dipion. Furthermore, at the $e^+e^-$ energy in the region of the possible higher peak near the threshold of $B_1 \bar B^*$ the same dipion decay of $B_1$ can produce $I^G = 1^-$ pairs $B^* \bar B + c.c.$ for which an isovector resonance $W_{b1}$ with $J^P=1^+$ is also expected at the threshold. 

In summary. We discuss the possibility that an $S$-wave production of the heavy meson pairs $B_1(5721) \bar B + c.c.$ takes place near their threshold in $e^+e^-$ annihilation at the energy of the $\Upsilon(6S)$ peak. This would lead to the enhanced yield in the channel $Z_b(10610)\pi$ due to the triangle singularity mechanism illustrated in Fig.~1 with a subsequent production in the decays of the $Z_b$ resonance of final states with ortho- and para- bottomonium, $e^+e^- \to \Upsilon(nS) \pi \pi$ and $e^+e^- \to h_b(kP) \pi \pi$. If this mechanism is dominant any non-resonant background in these final channels, not associated with the $Z_b(10610)$, should be strongly suppressed. Furthermore, the sub dominant decay of the $B_1$ meson, $B_1 \to B \pi \pi$, may provide, due to a similar triangle singularity, a gateway for studies of an isovector molecular resonance $W_{b0}$ expected at the threshold of $B \bar B$. The discussed picture also suggests that there may be a similar structure in the $e^+e^-$ annihilation at energy near the threshold for the pairs $B_1 \bar B^*$ and $B_2 \bar B^*$, i.e. in the vicinity of 11.06\,GeV.  The apparent deficiency of the discussed mechanism is the lack of a quantitative explanation for production of $B_1(5721) \bar B + c.c.$ in the $S$ wave which production breaks HQSS. However the assumption of existence of such mechanism leads to a number of distinctive features in the $e^+e^-$ annihilation near the c.m. energy 11.00\,GeV that can be studied in experiments. We thus believe that testing those features makes sense and may shed some light on yet poorly understood dynamics of heavy mesons near their threshold. 

We thank Alexei Garmash and Roman Mizuk for stimulating discussions.
The work of M.B.V. is supported in part by U.S. Department of Energy Grant No.\ DE-SC0011842.


\begin{thebibliography}{99}
\bibitem{belle1501} 
  D.~Santel {\it et al.} [Belle Collaboration],
  Phys.\ Rev.\ D {\bf 93}, no. 1, 011101 (2016)
  doi:10.1103/PhysRevD.93.011101
  [arXiv:1501.01137 [hep-ex]].
  
\bibitem{belle1508} 
  A.~Abdesselam {\it et al.} [Belle Collaboration],
  arXiv:1508.06562 [hep-ex].
  
\bibitem{bellez} 
  A.~Bondar {\it et al.}  [Belle Collaboration],
  Phys.\ Rev.\ Lett.\  {\bf 108}, 122001 (2012)
  [arXiv:1110.2251 [hep-ex]]. 
  
\bibitem{bgmmv}
  A.~E.~Bondar, A.~Garmash, A.~I.~Milstein, R.~Mizuk, M.~B.~Voloshin,
  Phys.\ Rev.\  {\bf D84}, 054010 (2011).
  [arXiv:1105.4473 [hep-ph]].
  
\bibitem{lt} 
  P.~V.~Landshoff and S.~B.~Treiman,
  Phys.\ Rev.\  {\bf 127}, 649 (1962).

\bibitem{ll} 
  X.~H.~Liu and G.~Li,
  Phys.\ Rev.\ D {\bf 88}, 014013 (2013)
  doi:10.1103/PhysRevD.88.014013
  [arXiv:1306.1384 [hep-ph]].
  
\bibitem{lv} 
  X.~Li and M.~B.~Voloshin,
  Phys.\ Rev.\ D {\bf 88}, no. 3, 034012 (2013)
  doi:10.1103/PhysRevD.88.034012
  [arXiv:1307.1072 [hep-ph]].
  
\bibitem{mv12} 
  M.~B.~Voloshin,
  Phys.\ Rev.\ D {\bf 84}, 031502 (2011)
  doi:10.1103/PhysRevD.84.031502
  [arXiv:1105.5829 [hep-ph]].
  
  
\bibitem{ding}
  G.~J.~Ding,
  Phys.\ Rev.\ D {\bf 79}, 014001 (2009)
  doi:10.1103/PhysRevD.79.014001
  [arXiv:0809.4818 [hep-ph]].
  
\bibitem{lwdz} 
  M.~T.~Li, W.~L.~Wang, Y.~B.~Dong and Z.~Y.~Zhang,
  arXiv:1303.4140 [nucl-th].
  
\bibitem{whz} 
  Q.~Wang, C.~Hanhart and Q.~Zhao,
  Phys.\ Rev.\ Lett.\  {\bf 111}, no. 13, 132003 (2013)
  doi:10.1103/PhysRevLett.111.132003
  [arXiv:1303.6355 [hep-ph]].
  
\bibitem{cwghmz} 
  M.~Cleven, Q.~Wang, F.~K.~Guo, C.~Hanhart, U.~G.~Meißner and Q.~Zhao,
  Phys.\ Rev.\ D {\bf 90}, no. 7, 074039 (2014)
  doi:10.1103/PhysRevD.90.074039
  [arXiv:1310.2190 [hep-ph]].
  
\bibitem{lhcb} 
  R.~Aaij {\it et al.} [LHCb Collaboration],
  JHEP {\bf 1504}, 024 (2015)
  doi:10.1007/JHEP04(2015)024
  [arXiv:1502.02638 [hep-ex]].
  
\bibitem{belle05} 
  K.~Abe {\it et al.} [Belle Collaboration],
  Phys.\ Rev.\ Lett.\  {\bf 94}, 221805 (2005)
  doi:10.1103/PhysRevLett.94.221805
  [hep-ex/0410091].
  
\bibitem{dv} 
  S.~Dubynskiy and M.~B.~Voloshin,
  Phys.\ Rev.\ D {\bf 74}, 094017 (2006)
  doi:10.1103/PhysRevD.74.094017
  [hep-ph/0609302].
  
\bibitem{belle1512} 
  A.~Garmash {\it et al.} [Belle Collaboration],
  arXiv:1512.07419 [hep-ex].
  
\bibitem{pdg} 
  K.~A.~Olive {\it et al.}  [Particle Data Group Collaboration],
  Chin.\ Phys.\ C {\bf 38}, 090001 (2014).
  
\end{thebibliography}
\end{document}